\documentclass[aps,prd,preprint,eqsecnum]{revtex4}

\usepackage{amsmath,amssymb,amsfonts,verbatim}
\usepackage[dvips]{graphicx}

\def\beqn{\begin{eqnarray}}
\def\eeqn{\end{eqnarray}}
\def\nn{\nonumber}
\def\spa#1.#2{\langle#1#2\rangle}
\def\spb#1.#2{[#1#2]}

\def\spRa#1.#2{\langle#1#2]}
\def\spRb#1.#2{[#1#2\rangle}
\def\lp#1.#2{p_#1 \cdot p_#2}
\def\lphat#1{\widehat P \cdot p_{\widehat{#1}}\,}
\def\lphatR#1{\widehat P_R \cdot p_{\widehat{#1}}\,}
\def\lphatRb#1{\widehat P_R \cdot p_{#1}\,}
\def\lphatS#1.#2{p_{\widehat{#1}} \cdot p_#2\,}
\def\lphatT#1.#2{p_#1 \cdot p_{\widehat{#2}}\,}

\def\ang#1{|#1\rangle}
\def\bra#1{|#1]}
\def\bang#1{\left\langle#1\right|}
\def\bbra#1{[#1|}
\newcommand\ab[3]{\langle#1|#2|#3]}
\newcommand\ba[3]{[#1|#2|#3\rangle}

\def\pslash{\hbox{$\slash \hskip -.22cm p$}}
\def\Pslash{\hbox{$\slash \hskip -.24cm P$}}
\def\epslash{\hbox{$\slash \hskip -.20cm \epsilon$}}
\def\wh#1{\widehat #1}

\begin{document}
\title{Massive Quark-Gluon Scattering Amplitudes at Tree Level}
\author{Anthony Hall}
\affiliation{{} Department of Physics and Astronomy, UCLA\\
\hbox{Los Angeles, CA 90095--1547, USA}
\\{\tt anthall@physics.ucla.edu}
}

\date{10/08/07}

\begin{abstract}
Results for four-, five-, and six-parton tree amplitudes for massive
quark-antiquark scattering with gluons are calculated using the
recursion relations of Britto, Cachazo, Feng, and Witten.  The
required diagrams are generated using shifts of the momenta of a pair
of massless legs to complex values.  Checks verifying the calculations
are described, and a simple formula for the shifted spinors of an internal 
gluon is presented.

\end{abstract}

\maketitle

\section{Introduction}

Stimulated by Witten's introduction of a string theory on twistor
space dual to perturbative gauge theory \cite{Witten}, significant
recent progress has been made in the calculation and understanding of the 
structure of gauge-theory
amplitudes.  By revealing the maximally helicity violating amplitudes
to be primitive vertices from which an amplitude can be computed,
Feynman diagram calculations can be replaced by the MHV rules of
Cachazo, Svr\v cek, and Witten \cite{CSW} relating the desired tree
amplitude with products of propagators and on-shell MHV amplitudes
that possess fewer legs.  Britto, Cachazo, Feng, and Witten (BCFW)
\cite{BCFrec,BCFWproof} later proved that a tree amplitude is given by
the sum of products of a propagator and two on-shell amplitudes with
fewer legs and shifted, complex momenta.  This recursion yields very
compact expressions for an amplitude.  Recursive methods for the
calculation of leading-order QCD processes have existed for many years
\cite{BGRecursion, LaterRecursion}, initially formulated by Berends
and Giele.  In contrast to the Berends-Giele recursion relation which 
is based on off-shell vertices, the building blocks 
for the new methods are on-shell tree amplitudes.  On-shell methods have been
previously used at loop level via the unitarity method~\cite{Unitarity}.

The BCFW recursion, while initially derived for pure gauge boson tree
amplitudes, has been generalized to include massive particles with
spin through the work of Badger, Glover, Khoze, and Svr\v cek
\cite{BGKS}.  Thus the new recursive methods can be brought to bear on
a variety of phenomenologically interesting problems to be encountered
at the Large Hadron Collider such as top quark, vector boson, and
possible supersymmetric processes.  The multigluon scattering with a
single external massive gauge boson or Higgs boson has been calculated
\cite{Dixonhiggs, Badgerhiggs1, Bernvector, Badgerhiggs2} as well as
processes with a pair of massive scalar particles
\cite{BGKS,Fordescalars}.  The supersymmetry Ward identities of
supersymmetric QCD
have been applied to these scalar interactions, yielding the quark
amplitudes for certain helicities \cite{Schwinnward}.  Besides
tree-level gauge theory
\cite{Luo1,Luo2,Britto,BGKS,BGK,Fordescalars,Quigley,Ferrario,Stirling,Dinsdale,Duhr,Draggiotis,deFlorian1,deFlorian2},
the BCFW recursion has been utilized for one-loop QCD calculations
\cite{Bernloop1,Bernfiniteloops,Bernboot,Berncoeff,FordeMHVloop,Bergergeneralloop,BergerMHVloop,Bergerscalar}
and gravity \cite{Bedford,Cachazo,Bjerrum,Brandhuber,Benincasa}.

The present paper is concerned with the tree-level scattering of a
massive fermion pair with massless gauge bosons.  Results for
$\bar{q}q \rightarrow ggg$ were presented by Ozeren and Stirling, but
these authors report that difficulties are encountered with the BCFW
recursion so that Feynman diagrams were necessary when gluons are
exchanged for certain helicity and spin configurations~\cite{Stirling}.  Here
we remedy this and present an explicit formula for the spinors
associated with the massless gluon exchanged within a diagram.  In
this paper, the massive quark-gluon tree amplitudes with six and fewer
partons are presented.  The necessary recursion diagrams are found
solely via the BCFW method with spinor shifts on two of the external
massless gluon legs.  Recent work by Schwinn and Weinzierl
\cite{Schwinnborn} also deals with scattering of massive quarks which
also avoid the potential difficulties.  In that reference a formula is
given for the scattering of a pair of massive quarks with an arbitrary
number of gluons in a specific helicity configuration, one negative-helicity 
gluon and the rest positive.  Here we provide all the
remaining helicity configurations up to amplitudes with four external gluons.

\section{Review of the Spinor-Helicity Formalism}

The tree amplitudes presented here are the color-ordered partial
amplitudes $A_n$, containing the kinematic data, from which the full
amplitude $M_n$ with color information is determined by the color
decomposition \cite{Dixon},
\beqn
	M_n(k_i,\lambda_i,a_i) = g^{n-2} \sum_{\sigma \in S_{n-2}} 
	{(T^{a_{\sigma(3)}} \cdots T^{a_{\sigma(n)}})_{i_2}}^{\bar{j}_1}
	A_n(\bar{1}^{\lambda_1}_q,2^{\lambda_2}_q,{\sigma(3^{\lambda_3})}_g,
		\cdots {\sigma(n^{\lambda_n})}_g) .
\eeqn
The notation $k^{\lambda_k}$ labels the $k$-th particle with spin
$\lambda_k$, $S_{n-2}$ is the group of permutations on $n-2$ labels,
and the quark colors are $i_2$ and $\bar{j}_1$ for the quark labelled
$2$ and antiquark $1$.  The color generators are normalized as ${\rm tr} \,
T^a T^b=\delta^{a b}$.

In the spinor helicity formalism, spinors for the particle with
momentum $p_i$ are denoted
\beqn
	u^+(p_i) \equiv \ang{i}, & \overline{u^+}(p_i) \equiv \bbra{i},\nn & \\
	u^-(p_i) \equiv \bra{i}, & \overline{u^-}(p_i) \equiv \bang{i}.
\eeqn
For massive spinors, the same angle and square brackets denote spin with 
respect to a fixed axis
rather than helicity.  These massive spinors have four components and
satisfy the Dirac equation, and they are interpreted as Weyl spinors
only in the massless limit.

Spinor products take the form
\beqn
\overline{u^-}(p_i) u^+(p_j) = \spa i.j, & 
\overline{u^+}(p_i) u^-(p_j) = \spb i.j, & \nn \\ \label{spRdef}
\overline{u^+}(p_i) u^+(p_j) = \spRb i.j, &
\overline{u^-}(p_i) u^-(p_j) = \spRa i.j .
\eeqn
We use the formalism of Kleiss and Stirling \cite{Kleiss} for the construction of massive 
spinors, which yield 
\beqn
\spa i.j &=& 
\frac{(p_j \cdot k_0) (p_i \cdot k_1) - (p_i \cdot k_0) (p_j \cdot k_1)
-i\epsilon_{\mu \nu \rho \sigma}k_0^\mu p_i^\nu p_j^\rho k_1^\sigma}
		{\sqrt{(p_i \cdot k_0) (p_j \cdot k_0)}} , \nn \\
\spb i.j &=& 
\frac{(p_i \cdot k_0) (p_j \cdot k_1) - (p_j \cdot k_0) (p_i \cdot k_1)
	-i\epsilon_{\mu \nu \rho \sigma}k_0^\mu p_i^\nu p_j^\rho k_1^\sigma}
		{\sqrt{(p_i \cdot k_0) (p_j \cdot k_0)}} , \nn \\
\spRb i.j &=& \frac{m_i (p_j \cdot k_0) + m_j (p_i \cdot k_0)}
		{\sqrt{(p_i \cdot k_0) (p_j \cdot k_0)}} ,	
\eeqn
where $k_0$ and $k_1$ are vectors satisfying $k_0^2=0$, $k_1^2=-1$,
and $k_0 \cdot k_1=0$.  The vector $k_0$ corresponds to the axis of
spin quantization for the massive fermions.  Implicit in these spinor-product 
formulas is their analytic continuation under crossing
symmetry.  A past-pointing momentum $p_i$ is replaced with $-p_i$
while each formula acquires an overall factor of $i$, and $m_i$
becomes $-m_i$ ($m > 0$ for all legs).

The final spinor product required is $\langle i|\pslash|j]$, whose formula is determined 
by the spin-sum rule for a particle with momentum $p$, $\ang{p}\bbra{p}+\bra{p}\bang{p} = \pslash+m_p$, so that
\beqn
	\langle i|\pslash|j] = \spa i.p \spb p.j + \spRa i.p \spRa p.j - m_p \spRa i.j .
\eeqn
If  $p$ corresponds to an antiparticle, $m_p \mapsto -m_p$. The shorthand $\langle i|j|k]$ is used for this spinor product.  
Under complex conjugation, the spinor products transform as
\beqn
	\spa{a}.{b}^\ast &=& \spb{b}.{a}, \\ \nn
	\spRb{a}.{b}^\ast &=& \spRa{a}.{b}, \\ \nn
	\ab{a}{k}{b}^\ast &=& \ba{a}{k}{b}.
\eeqn

The Schouten identity for two-component spinors is 
\beqn	\label{schouten}
	\ang{a}\spa b.c + \ang{c}\spa a.b + \ang{b}\spa c.a = 0 ,
\eeqn
and we have spinor identities for massive legs $i$ and $j$,
\beqn	\label{massiveschouten}
\spb a.b \Big(\spa i.a \spRa b.j - \spa i.b \spRa a.j \Big) + 
\spa a.b \Big(\spRa i.a \spb b.j - \spRa i.b \spb a.j \Big) = 
2\lp a.b \spRa i.j,
\eeqn
\beqn	\label{massiveschouten2}
	m_i \spa b.c + \spRb i.c \spa i.b + \spRb i.b \spa c.i = 0,
\eeqn
and
\beqn
	\label{spR}
	\spRb i.a \spRb j.b = \spRb i.b \spRb j.a .  
\eeqn 
Complex conjugation of these identities yields their square-bracket
counterparts.

For helicity-labelling conventions we consider all the gluons to be
outgoing so that a helicity label on a gluon denotes its helicity
leaving a vertex.  The massive quarks are both considered to be
incoming.  Thus momentum conservation means $p_1+p_2 =
p_3+\ldots+p_n$.  Our fermion spin labels are therefore opposite to the 
helicity labels for outgoing particles in the massless limit.

Simple relations exist between $n$-point amplitudes
with different helicities and spins.  Complex conjugation of an
amplitude flips the spin of every particle,
\beqn
A_n\left(\bar{1}^{s_1}_q,2^{s_2}_q, 3^{s_3}_g, \ldots, n^{s_n}_g \right)^\ast
	=
A_n\left(\bar{1}^{-s_1}_q,2^{-s_2}_q, 3^{-s_3}_g, \ldots, n^{-s_n}_g \right),
\label{ComplexConjugate}
\eeqn
and the spin of an individual quark may be flipped by replacing
bracket spinors with angle spinors, or vice versa, so we have
\beqn
A_n\left(\bar{1}^{\pm}_q, 2^{s_2}_q, 3^{s_3}_g, \ldots, n^{s_n}_g \right) 
\Big|_{\hbox{\footnotesize$\genfrac{}{}{0cm}{0}{\bbra1 \mapsto \bang1}{\bang1 \mapsto \bbra1}$}}
	&=&
A_n\left(\bar{1}^{\mp}_q, 2^{s_2}_q, 3^{s_3}_g, \ldots, n^{s_n}_g \right),
\nn     \\ 
A_n\left(\bar{1}^{s_1}_q, 2^{\pm}_q, 3^{s_3}_g, \ldots, n^{s_n}_g \right)
  \Big|_{\hbox{\footnotesize $\genfrac{}{}{0cm}{0}{\bra2 \mapsto \ang2}{\ang2 \mapsto \bra2}$}}
        &=&
 A_n\left(\bar{1}^{s_1}_q, 2^{\mp}_q, 3^{s_3}_g, \ldots, n^{s_n}_g \right).
\label{FlipQuark}
\eeqn
Thus, for example, complex conjugation of the amplitude
$A_n\left(\bar{1}^{-}_q, 2^{-}_q, 3^{s_3}_g, \ldots, n^{s_n}_g
\right)$ followed by the replacements $\bang1 \mapsto \bbra1$ and
$\ang2 \mapsto \bra2$ flips the helicity of only the gluons, yielding
the amplitude $A_n\left(\bar{1}^{+}_q, 2^{+}_q, 3^{-s_3}_g, \ldots,
n^{-s_n}_g \right)$.  Explicitly following the terms $\spb1.a$ and
$\spRb1.a$ through this composition of maps we have
\beqn
\spb1.a &\buildrel{\ast}\over{\mapsto}& -\spa1.a \buildrel{\bang1 \mapsto \bbra1}\over{\longmapsto} 
-\spRb1.a , \\ \nn
\spRb1.a &\buildrel{\ast}\over{\mapsto}& \spRa1.a \buildrel{\bang1 \mapsto \bbra1}\over{\longmapsto} 
\spb1.a .
\eeqn
Every possible combination of spins can be obtained from the
amplitudes $A_n\left(\bar{1}^{-}_q, 2^{-}_q, 3^{s_3}_g, \ldots,
n^{s_n}_g \right)$ by a combination of complex conjugation and quark
spinor replacements.  Thus in presenting the results for the amplitudes
we need only present the $\bar 1_q^-, 2_q^-$ amplitudes and half of the
possible gluon helicity configurations.  
The spinor 
identities~(\ref{schouten}),~(\ref{massiveschouten}),~(\ref{massiveschouten2}),~and~(\ref{spR}) can 
only be used when the massive fermions have particular spins, so their 
use will invalidate the above relations between amplitudes.
We present the amplitudes in 
a form where no spin-dependent identities are applied in order to allow 
relations~(\ref{ComplexConjugate}) and~(\ref{FlipQuark}) to be used.

The amplitudes presented in ref.~\cite{Schwinnborn} use a different
spinor formalism which must be accounted for in comparison of the amplitudes.
 The spin quantization axis $k_0$ is chosen to be
one of the massless gluon legs, labelled as leg $q$, and massive
vectors are projected to the light cone with this leg $q$,
\beqn
k^{\flat} = k-\frac{k^2}{2 k\cdot q}q.
\eeqn
Massive spinors are defined as
\beqn
u^+(k) &=& \ang{k^{\flat}}+\frac{m}{\spb k^{\flat}.q} \bra{q}, \nn \\
u^-(k) &=& \bra{k^{\flat}}+\frac{m}{\spa q.{k^{\flat}}} \ang{q} ,
\eeqn
which leads to the spinor products
\beqn
\spa i.j &=& \spa i^{\flat}.{j^{\flat}}, \quad \spb i.j 
     = \spb i^{\flat}.{j^{\flat}}, \nn \\ 
\spRb i.j&=& m_i \frac{\spa q.j}{\spa q.i} + m_j \frac{\spb q.i}{\spb q.j}.
\eeqn

\section{The BCFW Recursion Formula}

The BCFW tree-level recursion formula follows from the basic complex
analytic properties of amplitudes in gauge theory.
Consider a tree-level color-ordered gauge-theory amplitude
$A_n\left(p_1,\ldots,p_n\right)$ with complex momenta obtained by
shifting a pair of spinors from the massless external legs $i$ and $j$,
\beqn \label{shift}
	\widehat{\ang{i}} \equiv \ang{i}+z\ang{j}, 
	&  \widehat{\bra{j}} \equiv \bra{j}-z\bra{i}.
\eeqn
Under this spinor shift, the momenta shift as 
\begin{eqnarray}
	\wh p_i  &=& \ang{i} \bbra{i}+z \ang{j} \bbra{i}, \nonumber \\
	\wh p_j &=&  \ang{j} \bbra{j}-z \ang{j} \bbra{i}.
\end{eqnarray}
To be concise we denote this choice of shifted legs ``$\spRa i.j$",
but from context this should not be confused with the spinor product
of eq.~(\ref{spRdef}).

\begin{figure}
\centering
\includegraphics{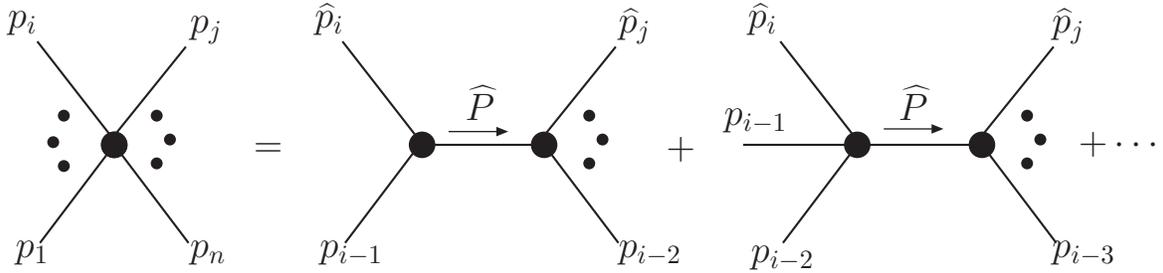}
\caption{An illustration of the recursion formula for cases where the
shifted external legs $p_i$ and $p_j$ are chosen to be adjacent with
$i<j$.  The sum over the internal particle's helicity is implicit.  We
denote the BCFW diagrams, from left to right, as $\mathcal D_1$,
$\mathcal D_2$, $\ldots$
\label{bcfwfigure}}
\end{figure}

This BCFW shift keeps all momenta on shell, and total momentum is still 
conserved,
	\begin{eqnarray*}
		\wh p_i + \wh p_j = p_i+p_j.
	\end{eqnarray*}
The amplitude is now a meromorphic function $A_n(z)$ with simple poles
in $z$ when the sum of adjacent particles' momenta, denoted $\widehat
P$, is $z$-dependent and on shell,
\begin{equation}
0 = \wh P^2 - m^2 = P^2 - m^2 + z \ab{j}{P}{i} .
\end{equation}
Provided  that $A_n(z)$ vanishes as $z \rightarrow \infty$, we have $\int \! A(z)/z = 0$.  
Applying Cauchy's theorem to $\int \! A(z)/z$
yields residues for $z=0$ (the desired unshifted amplitude) and the values 
of $z$ for which $\widehat P^2-m^2 = 0$.
The factorization 
of amplitudes at these poles yields the BCFW formula \cite{BCFWproof}
\beqn
	A_n\left(p_1,\ldots,p_n\right) = 
	\sum_{\rm partitions} \sum_{h} 
	\widehat{A}_{\rm left}\left(p_L,\ldots,\widehat P^h\right)
	\times \frac{1}{P^2-m_P^2} \times \widehat{A}_{\rm right}\left(-\widehat P^{-h},\ldots,p_R\right), 
\hskip20pt
\eeqn
as illustrated in fig.~\ref{bcfwfigure}.
Each term appearing in the recursion formula is one of the residues of
$A(z)/z$.  The momentum $\widehat P$ is the shifted total momentum
leaving from the right amplitude's external legs, and the recursion
formula sums over the helicity of the internal state with momentum
$P$.  The sub-amplitudes $\widehat{A}_{\rm{left/right}}$ are to be
evaluated with the shifted spinors at the value of $z$ for which
$\widehat P^2-m^2 = 0$.
%
%
%

Valid recursion relations are obtained only for certain helicities of
the shifted external legs, where $A_n(z)$ vanishes as $z \rightarrow
\infty$.
Ref.~\cite{Schwinnborn} derives the valid choices of legs to shift
for both massive and massless particles.  We shift only massless
gluons in this paper, so it suffices to note that the recursion is
valid for the ``$\spRa i.j$" shift in eq.~(\ref{shift}) provided that
the shifted legs $i$ and $j$ have helicities~\cite{BCFWproof}
\beqn
	(h_i,h_j) = (+,+),& (+,-),& \rm{or~} (-,-).
\eeqn

By applying the spin-sum rule to combine the spin states of an
internal fermion, rather than considering spin states separately, a
term with a fermionic-particle pole in the recursion formula can
instead be written as
\beqn
\widehat{A}_{\rm left}\left(p_L,\ldots,\widehat P^\ast\right)
\times \frac{\widehat\Pslash+m}{P^2-m_P^2} \times 
\widehat{A}_{\rm right}\left(-\widehat P^\ast,\ldots,p_R\right),
\eeqn
where $P^\ast$ denotes that the internal spinors have been stripped
off the amplitudes.  We use this method, from ref.~\cite{BGK}, to
calculate the BCFW diagrams where an internal fermion is exchanged
between trees.  
An amplitude which is to be fed into the recursion should not 
be simplified in advance using any identities which depend on the spin of the 
massive legs.  Doing so violates the rule that both spin states of an internal 
fermion should be summed over, not just the spin for which a particular 
identity holds.

Consider the spinors $\widehat{\ang{P}}$ and $\widehat{\bra{P}}$ which
appear in the recursion diagrams for the exchange of a massless
particle. Because the internal particle appears with opposite helicity
at each sub-amplitude, the overall amplitude will have degree zero in
the $\wh P$ spinors.  Thus identities such as $\langle a \wh P \rangle
[\wh P b] = \ab{a}{\wh P}{b}$ may be used to eliminate the individual
spinors.  However, to facilitate computer programmable calculations of
the diagrams, it is convenient to have a direct substitution for
$\widehat{\ang{P}}$ and $ \widehat{\bra{P}}$.  Moreover, an explicit
formula for the $\wh P$ spinors solves the problem of dealing with
massive spinor products such as $\spRb1.{\wh P}$ encountered in 
ref.~\cite{Stirling}.  To achieve this, we use the two-dimensionality of
massless spinors to write the $\wh P$ spinors in terms of the shifted
legs $i$ and $j$ (the shift is ``$\spRa i.j$"),
\beqn
	\widehat{\ang{P}} = \alpha \ang{i} + \beta \ang{j}, & & 
	\widehat{\bra{P}} = \gamma \bra{i} + \delta \bra{j}.
\eeqn
Then by using the identities
\beqn
	\ab{\wh P}{\wh P}{i} &=& \ab{j}{\wh P}{\wh P} = 0, \nn \\
	\ab{j}{\wh P}{i} &=& \ab{j}{P}{i},
\eeqn
we find 
\beqn
	\wh \Pslash &=& \widehat{\ang{P}} \widehat{\bbra{P}} = 
	\frac{\ab{j}{P}{i}}{2\lp i.j} 
	\left(\ang{i}-\frac{2 p_i \cdot P}{\ab{j}{P}{i}} \ang{j}\right)
	\left(\frac{2 p_j \cdot P}{\ab{j}{P}{i}} \bbra{i}-\bbra{j}\right)\nn \\ 
&=& \frac{1}{\ab{j}{P}{i}} \left( P\bra{i} \right) \left( \bang{j}\! P \right),
\eeqn
providing a simple formula for the shifted spinors of an internal gluon.
The overall prefactor may be associated with either $\widehat{\ang{P}}$
or $\widehat{\bra{P}}$.  This expression for $\wh P$ is valid no
matter on which side, left or right, the legs $i$ and $j$ lie.

\section{$\bar{q}qgg$ Amplitudes}

\begin{figure}
\centering
\includegraphics{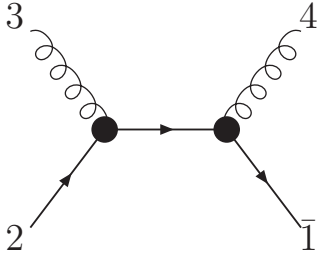}
\caption{
	\label{2gshift34}
The single BCFW diagram contributing to all the four-point amplitudes.}
\end{figure}

As input for the single BCFW diagram of fig.~\ref{2gshift34}
contributing in the amplitudes with two gluons we use the polarization
vectors of positive- and negative-helicity gluons with momentum $p$,
\beqn
\epslash^+(p,q) &=& \frac{\sqrt2}{\spa{q}.{p}}
		\Big(\ang{q}\bbra{p}+\bra{p}\!\bang{q}\Big), \nn \\
\epslash^-(p,q) &=& \frac{\sqrt2}{\spb{q}.{p}}
		\Big(\ang{p}\bbra{q}+\bra{q}\!\bang{p}\Big),
\eeqn
with the three-vertex factor $\frac{1}{\sqrt2}$.  The spinors $q$
refer to an arbitrary null vector.  Different choices of reference
spinors lead to gauge-equivalent polarization vectors.

The four-point amplitudes with opposite-helicity gluons are
\beqn
A_4\left(\bar{1}^{-}_q,2^{-}_q,3^{+}_g,4^{-}_g\right) = 
\ab{4}{1}{3} \frac{[13] \spRa4.2+\spRb1.4 [32]}{4 \lp3.4  \lp1.4}
\eeqn
and
\beqn
A_4\left(\bar{1}^{-}_q,2^{-}_q,3^{-}_g,4^{+}_g\right) = 
\ba{4}{1}{3} \frac{\spRb1.3 [42]+[14] \spRa3.2}{4 \lp3.4  \lp1.4}.
\eeqn

%
The four-point amplitudes with identical-helicity gluons are given by
\beqn \label{mm}
A_4\left(\bar{1}^{-}_q,2^{-}_q,3^{-}_g,4^{-}_g\right) = m \spa4.3
\frac{[13][42]-[14][32]} {[34]^2 2\lp2.3 }
\eeqn
and 
\beqn
\label{pp}
A_4\left(\bar{1}^{-}_q,2^{-}_q,3^{+}_g,4^{+}_g\right) & = & 
m [34] \frac{ \spRb1.3 \spRa4.2 - \spRb1.4 \spRa3.2 }
 { \spa3.4^2 2\lp2.3 } .
\eeqn
Using eq.~(\ref{spR}) this amplitude vanishes but only for this
specific choice of quark spins. To obtain the amplitudes with the
other spins, one starts from eq.~(\ref{pp}) and applies
eqs.~(\ref{ComplexConjugate}) and~ (\ref{FlipQuark}).
One must include this amplitude in this form as
input to higher-point recursion diagrams with internal quarks, since
the internal quark is summed over all spins and not only the spin for
which this amplitude happens to vanish.

Likewise note that the Schouten identity could be applied in
eq.~(\ref{mm}) to yield $[13][42]-[14][32] = [12][43]$, thus
cancelling the unphysical double pole $[34]^2$ in the denominator.
However this identity can be applied only for the particular quark
spins chosen, so we leave the expression as is to avoid invalidating
the next level of recursion.
%

\section{$\bar{q}qggg$ Amplitudes}

We now write the five-point amplitudes beginning with
$A_5\left(\bar{1}^{-}_q,2^{-}_q,3^{-}_g,4^{-}_g,5^{-}_g\right)$
calculated from the shift choice ``$\spRa4.5$", with the result
\beqn
A_5\left(\bar{1}^{-}_q,2^{-}_q,3^{-}_g,4^{-}_g,5^{-}_g\right) & = & 
m \Big( \spa3.4+\frac{2\lp1.5}{\ba{4}{1}{5}} \spa3.5 \Big) \hskip 210pt
 \\ && \times \nn
\Bigg[ \frac{m \spRb1.5 [43] [42] - [14] \ba{3}{1}{5} [42]
  + [14] \ba{4}{1}{5} [32]}
 {2\lp2.3 [34]^2 [45] 2\lp1.5} \Bigg].
\eeqn
%
%
\begin{figure}
\centering
\includegraphics{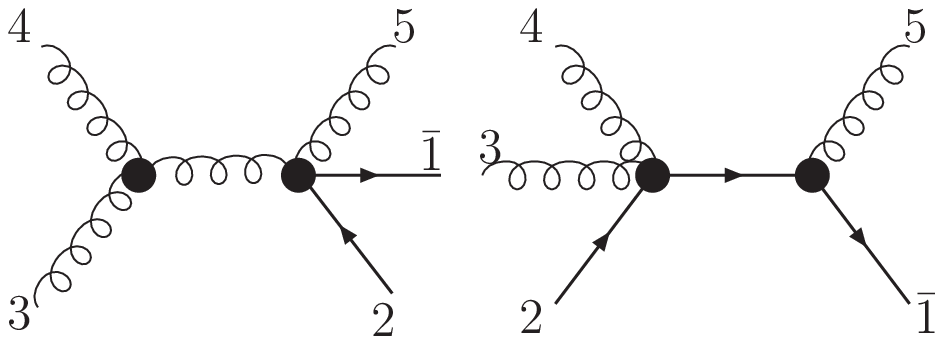}
\caption{\label{3gshift45}
The BCFW recursion diagrams required for the five-point
amplitudes under the shifts ``$\spRa4.5$" or ``$\spRa5.4$".  These
diagrams are relevant for the gluon helicities $(---)$ and $(+--)$.  }
\end{figure}

With the shift $``\spRa4.5"$ we find
\beqn
\lefteqn{A_5\left(\bar{1}^{-}_q,2^{-}_q,3^{+}_g,4^{-}_g,5^{-}_g\right) = 
\Big(\ba{3}{2}{4}+\frac{2\lp1.5}{\ba{4}{1}{5}}\ba{3}{2}{5}\Big) }
\\ && \times \nn
\Bigg[
\frac{
\big(-[14]\ba{3}{1}{5} + m \spRb1.5 [43] \big)
\big(\spRa4.2+\frac{2\lp1.5}{\ba{4}{1}{5}}\spRa5.2\big) +
\big(-\spRb1.5 2\lp2.3 +m[14]\spa5.4 \big) [32]}
{8 \lp2.3 \big(\lp3.4+\frac{\lp1.5}{\ba{4}{1}{5}}\ba{4}{3}{5}\big)
 \lp1.5 [45]} 
\Bigg] \hskip 22pt \\ && \nn 
\null +
\frac{m \spa4.5^3}{\spa4.3 (p_1+p_2)^4 (2\lp1.5+\frac{\spa3.4}{\spa3.5}\ab{5}{1}{4})}
\times \Bigg[[1 \wh P_R][\wh5 2]-[1 \wh5][\wh P_R 2] \Bigg].
\eeqn
For the second term, which arises from the first BCFW diagram in 
fig.~\ref{3gshift45}, we have $z = -\frac{\spa3.4}{\spa3.5}$ 
so that
\beqn
\ang{\wh4} &=& \ang4 - \frac{\spa3.4}{\spa3.5} \ang5
		= \frac{\spa4.5}{\spa3.5} \ang3 ,
		 \nn \\
\bra{\wh5} &=& \bra5 + \frac{\spa3.4}{\spa3.5}\bra4 , \nn \\
\wh P_R &=& p_3 + p_{\wh4} = \ang3 \bbra3 + \ang{\wh4} \bbra4
		 = \ang3 \Big(\bbra3 + \frac{\spa4.5}{\spa3.5}\bbra4\Big),
\eeqn
where we have used the Schouten identity.
Thus, for example, $[a \wh P_R] = [a3] + \frac{\spa4.5}{\spa3.5} [a4]$.  
We use this notation throughout the paper, indicating the spinors to be substituted 
for the internal massless momenta $\wh P$ or $\wh P_R$ by explicitly writing 
$\wh \Pslash$ as a bi-spinor.
%
%
%
\begin{figure}
\centering
\includegraphics{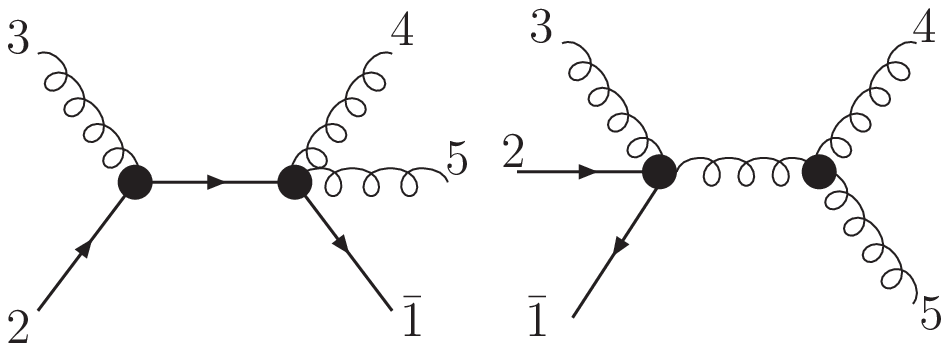}
\caption{\label{3gshift34}
The BCFW recursion diagrams required for the five-point
amplitudes under the shifts ``$\spRa3.4$" or ``$\spRa4.3$".  These
diagrams are relevant for the gluon helicities $(-+-)$ and $(--+)$.  }
\end{figure}

The shift ``$\spRa4.3$" leads to
\beqn
A_5\left(\bar{1}^{-}_q,2^{-}_q,3^{-}_g,4^{+}_g,5^{-}_g\right) = \ba{4}{1}{5}
\left[ 
\frac{m[14] \spa5.3 [42]+\spRb1.5 [42] \ba{4}{2}{3}-
      [14] \spRa3.2 \ba{4}{1}{5}}
{8 \lp1.5  \lp2.3  \left(\lp4.5+\frac{\lp2.3}{\ba{4}{2}{3}}
\ba{4}{5}{3}\right)  [34]} 
\right]  &&
\hskip 37.6pt 
\\ \null +
\frac{m \spa3.5^4}{\spa4.5 \spa3.4 (p_1+p_2)^4 (2\lp2.3+\frac{\spa4.5}{\spa3.5}\ab{3}{2}{4})}
\times \Bigg[[1\wh3] [\wh P 2]-[1 \wh P][\wh3 2] \Bigg], \nn
\eeqn
where, in the second term, which arises from the first BCFW diagram in 
fig.~\ref{3gshift34}, $z = -\frac{\spa4.5}{\spa3.5}$, which means
\beqn \label{mpmshift}
\bra{\wh3} &=& \bra3 + \frac{\spa4.5}{\spa3.5}\bra4 , \nn \\
\wh P &=& p_{\wh4} + p_5 = \ang5 \Big(\frac{\spa3.4}{\spa3.5}\bbra4+\bbra5\Big).
\eeqn

The amplitude 
$A_5\left(\bar{1}^{-}_q,2^{-}_q,3^{-}_g,4^{-}_g,5^{+}_g\right)$ is 
calculated from the shift ``$\spRa4.3$",
\beqn
\lefteqn{A_5\left(\bar{1}^{-}_q,2^{-}_q,3^{-}_g,4^{-}_g,5^{+}_g\right) = 
\Big(\ab{4}{1}{5}+\frac{2\lp2.3}{\ab{3}{2}{4}}\ab{3}{1}{5}\Big) }
\\ & & \times
\left[
\frac{[15] (m\spa4.3 [42]+2\lp1.5\spRa3.2)+
  \big(\spRb1.4+\frac{2\lp2.3}{\ab{3}{2}{4}}\spRb1.3\big)
  \big(\ab{3}{2}{5} [42]+m[54]\spRa3.2\big)}
{8 \lp1.5 \lp2.3 (\lp4.5+\frac{\lp2.3}{\ab{3}{2}{4}}\ab{3}{5}{4} ) [34]}
\right] \hskip 42.3pt \nn
\\ &&  \null -
\frac{m \spa3.4^3}{\spa4.5 (p_1+p_2)^4 (2\lp2.3+\frac{\spa4.5}{\spa3.5}\ab{3}{2}{4})}
\times \Bigg[[1\wh3] [\wh P 2]-[1 \wh P][\wh3 2] \Bigg], \nn
\eeqn
where the shifted spinors in the second term are the same as in
eq.~(\ref{mpmshift}).
%

\section{$\bar{q}qgggg$ Amplitudes}

The six-point amplitudes with zero- and one-positive-helicity gluon are
presented first.
We have 
\beqn
\lefteqn{A_6\left(\bar{1}^{-}_q,2^{-}_q,3^{-}_g,4^{-}_g,5^{-}_g,6^{-}_g\right) = 
-m\Big(\spa3.4+\frac{2\lphat5}{\ba{4}{\wh P}{\wh5}}\spa3.{\wh5}\Big)
\times
\frac{1}{2\lp2.3 [43]^2 [45] 2\lphat5 [65] 2\lp1.6}} \nn
 \\ & &
\times
\Bigg[
(\spRb1.6 2\lphat5 + m [15] \spa6.5)
 m [43] [42] \\ & & \nn
\null \hskip .5 cm 
+ 
(-[15] \ba{4}{1}{6} + m\spRb1.6 [54] )
(\ba{3}{\wh P}{\wh5} [42] - \ba{4}{\wh P}{\wh5} [32] )
\Bigg], \hskip 154pt
\eeqn
where the shift ``$\spRa5.6$" is used, so that
\beqn
	\nn \ang{\wh5} &=& \ang5 + \frac{2\lp1.6}{\ba{5}{1}{6}} \ang6 \mbox{,} \quad 
	\bra{\wh6} = \bra6 - \frac{2\lp1.6}{\ba{5}{1}{6}} \bra5 \mbox{,}\\ \nn
	\ba{a}{\wh P}{\wh5} &=& \ba{a}{2-3-4}{\wh5}, \\ 
	\lphat5 &=& -\lp2.3-\lp2.4+\lp3.4 .
\eeqn
%
%
\begin{figure}
\centering
\includegraphics{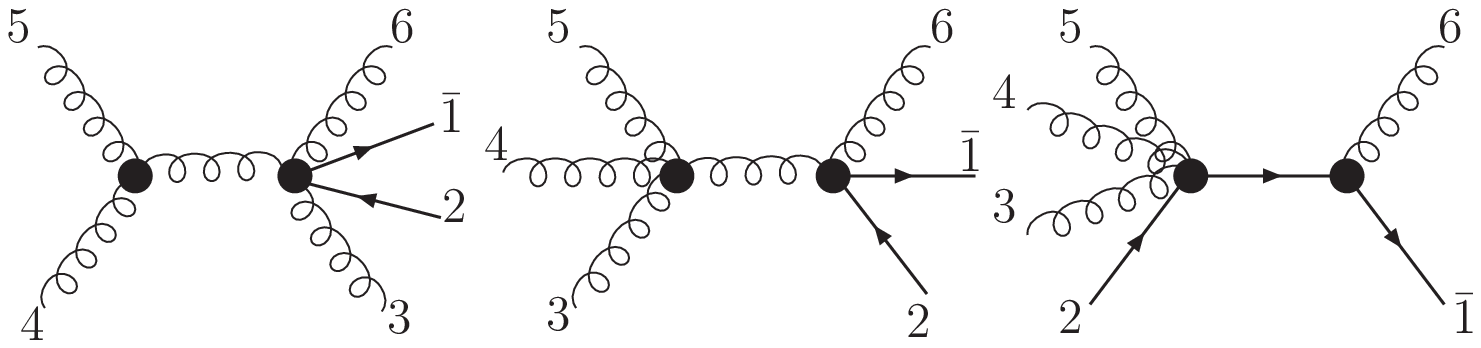}
\caption{
	\label{4gshift56}
The BCFW recursion diagrams required for the six-point amplitudes under the 
shifts ``$\spRa5.6$" or ``$\spRa6.5$".  This set of diagrams is relevant for 
the amplitudes with gluon helicities 
$(----\null)$, $(+---\null)$, $(-+--\null)$, $(-+-+\null)$, and $(-++-\null)$.}
\end{figure}
We also use the ``$\spRa5.6$" shift to obtain 
$A_6\left(\bar{1}^{-}_q,2^{-}_q,3^{+}_g,4^{-}_g,5^{-}_g,6^{-}_g\right)$.
This gives multiple
non-vanishing diagrams as shown in fig.~\ref{4gshift56} which we denote as
\beqn
A_6\left(\bar{1}^{-}_q,2^{-}_q,3^{+}_g,4^{-}_g,5^{-}_g,6^{-}_g\right)
= \mathcal D^{+---}_1 + \mathcal D^{+---}_2 + \mathcal D^{+---}_3,
\eeqn where $\mathcal D^{+---}_1=0$, and the second diagram under the
``$\spRa5.6$" shift yields
\beqn
\label{pmmm2}
\mathcal D^{+---}_2 =
-m\frac{\spa4.{\wh5}^3}{\ab{\wh5}{\wh P}{\wh6} \ab{3}{\wh P}{\wh6} \spa3.4   
p_{3,4,5}^2 2\lphatT1.6}
\times \Bigg[\ba{1}{\wh P_R}{6}[\wh6 2]+[1\wh6] \ab{6}{\wh P_R}{2} \Bigg],  \hskip63pt
\eeqn
where 
$z = -\frac{p_{3,4,5}^2}{\ab{6}{3+4}{5}}$ so that
\beqn \label{pmmmshift2}
	\ang{\wh5} &=& \ang5 - \frac{p_{3,4,5}^2}{\ab{6}{3+4}{5}} \ang6, \quad
	\bra{\wh6} = \bra6 + \frac{p_{3,4,5}^2}{\ab{6}{3+4}{5}} \bra5, \nn \\ 
	\ab{a}{\wh P}{\wh6} &=& -\ab{a}{(1+2)}{\wh6}, \nn \\ 
	\ab{6}{\wh P_R}{b} &=& \spa6.3[3b]+\spa6.4[4b]+\spa6.5[5b].
\eeqn
For the third diagram, $\mathcal D^{+---}_3$, we find
\beqn
\lefteqn{\mathcal D^{+---}_3 =
\Big(\ba{3}{2}{4}+\frac{2\lphat5}{\ba{4}{\wh P}{\wh5}}\ba{3}{2}{\wh5}\Big) } 
\label{pmmm3}
\\ && \times \nn
\frac{1}{2\lp2.3 \left(2\lp3.4+\frac{2\lphat5}{\ba{4}{\wh P}{\wh5}}\ba{4}{3}{\wh5}\right) 
	2\lphat5 [54] 2\lp1.6 [65]}  \\ & & \times \nn
\Bigg[
\left(\ba{3}{\wh P}{\wh5}\left(-[15] \ba{4}{1}{6}+m \spRb1.6 [54] \right) +
	m [43] (\spRb1.6 2\lphat5 + m[15] \spa6.5 ) \right) \\ &&  \nn 
\hskip .5 cm    \times
\Big( \spRa4.2+\frac{2\lphat5}{\ba{4}{\wh P}{\wh5}}\spRa{\wh5}.2 \Big) \\ && \nn
\hskip .5 cm \null - 
( \spRb1.6 2\lphat5+m[15] \spa6.5 ) 2\lp2.3 [32]  +
	m \left( -\ba{4}{1}{6} [15]+m [54] \spRb1.6 \right) \spa{\wh5}.4 [32]
\Bigg] \hskip 37pt \\ && \nn \null + 
m\frac{\spa4.{\wh5}^3}
	{[65] 2\lp1.6 \spa4.3 (p_1+p_2-p_{\wh6})^4 
	(2\lphatT1.5-2\lp5.6+\frac{\spa3.4}{\spa3.{\wh5}}(\ab{\wh5}{1}{4}-\ab{5}{\wh6}{4}))}
\\ && \times \nn
\Bigg[
\big(m\spRb1.6([53]+\frac{\spa4.{\wh5}}{\spa3.{\wh5}}[54])-
	[15](\ab{6}{1}{3}+\frac{\spa4.{\wh5}}{\spa3.{\wh5}}\ab{6}{1}{4})\big)
	\big([52]+\frac{\spa3.4}{\spa3.{\wh5}}[42]\big) \\ && \nn 
\hskip .5 cm \null -
\big(m\spRb1.6\frac{\spa3.4}{\spa3.{\wh5}}[54]-[15](\ab{6}{1}{5}+\frac{\spa3.4}{\spa3.{\wh5}}\ab{6}{1}{4})\big)
	\big([32]+\frac{\spa4.{\wh5}}{\spa3.{\wh5}}[42]\big)
\Bigg].
\eeqn
In $\mathcal D^{+---}_3$ we have,
\beqn \label{pmmmshift3}
	\ang{\wh5} &=& \ang5 + \frac{2\lp1.6}{\ba{5}{1}{6}} \ang6, \quad
	\bra{\wh6} = \bra6 - \frac{2\lp1.6}{\ba{5}{1}{6}} \bra5,\nn \\ \nn
	\lphat5 &=& -\lp2.3-\lp2.4+\lp3.4,  \\ 
	\ba{a}{\wh P}{\wh5} &=& \ba{a}{2-3-4}{\wh5}. 
\eeqn

From the shift $``\spRa5.6"$ we have 
\beqn
A_6\left(\bar{1}^{-}_q,2^{-}_q,3^{-}_g,4^{+}_g,5^{-}_g,6^{-}_g\right) = 
\mathcal D^{-+--}_1 + \mathcal D^{-+--}_2 + \mathcal D^{-+--}_3,
\eeqn
where the first diagram in fig.~\ref{4gshift56} is
\beqn
\lefteqn{\mathcal D^{-+--}_1 =
-m\left(\frac{\spa5.6}{\spa4.6}\right)^3 \times \frac{\spa3.4+\frac{2\lphatT1.6}{\ba{\wh P_R}{1}{6}}\spa3.6}
		{\spa5.4 2\lp2.3 \left(\frac{p_{3,4,5}^2}{\spa4.3}\right)^2 \frac{p_{4,5,6}^2}{\spa6.4}
		2\lphatT1.6} } \\ && \times \nn
\Bigg[
m\spRb1.6[\wh P_R 3][\wh P_R 2]-[1 \wh P_R] \ba{3}{1}{6} [\wh P_R 2] + [1 \wh P_R] \ba{\wh P_R}{1}{6}[32]
\Bigg], \hskip 140pt
\eeqn
with $\bra{\wh P_R} = \bra4 + \frac{\spa5.6}{\spa4.6} \bra5$.
The second diagram yields 
\beqn
\label{mpmm2}
\mathcal D^{-+--}_2 =
-m\frac{\spa3.{\wh5}^4} {\spa3.4 \spa4.{\wh5} \ab{\wh5}{\wh P}{\wh6} \ab{3}{\wh P}{\wh6} 
p_{3,4,5}^2 2\lphatT1.6}
\times \Bigg[\ba{1}{\wh P_R}{6}[\wh6 2]+[1\wh6] \ab{6}{\wh P_R}{2} \Bigg] \hskip 45pt
\eeqn
and the third diagram 
\beqn
\lefteqn{\mathcal D^{-+--}_3 = 
\frac{\ba{4}{\wh P}{\wh5}}{2\lphat5 2\lp2.3 
	\left(2\lphatT4.5+\frac{2\lp2.3}{\ba{4}{2}{3}}\ba{4}{\wh5}{3}\right)
	[34] 2\lp1.6 [65]} }  \label{mpmm1}
\\ && \nn
 \times
\Bigg[
\left(-[15] \ba{4}{1}{6}+m\spRb1.6[54]\right)
(\ba{4}{\wh P}{\wh5} \spRa3.2 + m\spa{\wh5}.3 [42]) \\ & & \nn
\hskip .5 cm \null +
(\spRb1.6 2\lphat5+m[15] \spa6.5) \ba{4}{2}{3} [42]
\Bigg] \hskip 240.5pt \\ && \nn
\null + 
\frac{m \spa3.{\wh5}^3}
	{\spa{\wh5}.4 \spa3.4 (p_1+p_2-p_{\wh6})^4 (2\lp2.3+\frac{\spa4.{\wh5}}{\spa3.{\wh5}}\ab{3}{2}{4})
	2\lp1.6 [65]} \\ && \nn \times
\Bigg[
\big(-[15](\ab{6}{1}{3}+\frac{\spa4.{\wh5}}{\spa3.{\wh5}}\ab{6}{1}{4})
	+m\spRb1.6 ([53]+\frac{\spa4.{\wh5}}{\spa3.{\wh5}}[54])\big) 
		\big([52]+\frac{\spa3.4}{\spa3.{\wh5}}[42]\big) \\ && \nn 
\hskip .5 cm \null -
\big(-[15](\ab{6}{1}{5}+\frac{\spa3.4}{\spa3.{\wh5}}\ab{6}{1}{4})+m\spRb1.6 \frac{\spa3.4}{\spa3.{\wh5}}[54]\big)
		\big([32]+\frac{\spa4.{\wh5}}{\spa3.{\wh5}}[42]\big)
\Bigg] , 
\eeqn
where the spinor shifts in $\mathcal D^{-+--}_2$ and $\mathcal D^{-+--}_3$ are the same as eqs.~(\ref{pmmmshift2}) and~(\ref{pmmmshift3}), respectively.

\begin{figure}
\centering
\includegraphics{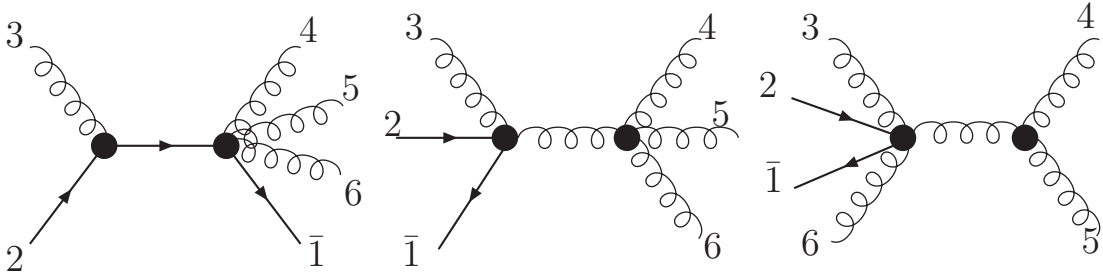}
\caption{
\label{4gshift34}
The BCFW recursion diagrams used for the six-point amplitudes under 
the shifts ``$\spRa3.4$" or ``$\spRa4.3$".  These diagrams are applied for 
the amplitudes with gluon helicities 
$(--+-\null)$, $(---+\null)$, and $(--++\null)$.}
\end{figure}

Choosing the shift ``$\spRa4.3$" leads to the diagrams in fig.~\ref{4gshift34},
\beqn
	A_6\left(\bar{1}^{-}_q,2^{-}_q,3^{-}_g,4^{-}_g,5^{+}_g,6^{-}_g\right) = 
	\mathcal D^{--+-}_1 + \mathcal D^{--+-}_2 + \mathcal D^{--+-}_3 ,
\eeqn
where we find
\beqn
\lefteqn{\mathcal D^{--+-}_1 = 
\frac{\ba{5}{1}{6}}
	{2\lp1.6 2\lphat4 
	\left(2\lp5.6+\frac{2\lphat4}{\ba{5}{\wh P}{\wh4}}\ba{5}{6}{\wh4}\right)
	[45] 2\lp2.3 [34] }} \label{mmpm1}
\\ & &  \nn
\times
\Bigg[
(\spRb1.6 \ba{5}{\wh P}{\wh4}+m[15] \spa6.{\wh4})
\left(\ba{5}{2}{3} [42]+m[54] \spRa3.2\right) \\ & & \nn
 \hskip .5 cm \null - 
[15] \ba{5}{1}{6} (2\lphat4 \spRa3.2+m\spa4.3 [42])
\Bigg]  \hskip 240pt \\ && \null +
m\frac{\spa{\wh4}.6^4}{\spa6.5 \spa{\wh4}.5 (p_1+p_2-p_{\wh3})^4 
	(2\lphat4+\frac{\spa5.6}{\spa{\wh4}.6}\ab{\wh4}{\wh P}{5}) [34] 2\lp2.3}  \nn \\ && \nn
\times
\Bigg[\Big([14]+\frac{\spa5.6}{\spa{\wh4}.6}[15]\Big)
	\Big( (\ab{3}{2}{6}+\frac{\spa{\wh4}.5}{\spa{\wh4}.6}\ab{3}{2}{5})[42]+
	m([64]+\frac{\spa{\wh4}.5}{\spa{\wh4}.6}[54])\spRa3.2\Big) \\ && \nn 
\hskip .5 cm \null -
	\Big([16]+\frac{\spa{\wh4}.5}{\spa{\wh4}.6}[15]\Big)
	\Big( (\ab{3}{2}{4}+\frac{\spa5.6}{\spa{\wh4}.6}\ab{3}{2}{5})[42]+
	m\frac{\spa5.6}{\spa{\wh4}.6}\spRa3.2\Big)
\Bigg].
\eeqn
The shifts for $\mathcal D^{--+-}_1$ are
\beqn \label{mmpmshift1}
	\ang{\wh4} &=& \ang4 + \frac{2\lp2.3}{\ba{4}{2}{3}} \ang3, \quad
	\bra{\wh3} = \bra3 - \frac{2\lp2.3}{\ba{4}{2}{3}} \bra4,\nn \\ \nn
	\ba{5}{\wh P}{\wh4} &=& \ba{5}{6-1}{\wh4},  \\
	\lphat4 &=& \lp1.5 + \lp1.6 - \lp5.6.
\eeqn
The second diagram, $\mathcal D^{--+-}_2$, is 
\beqn
\label{mmpm2}
\mathcal D^{--+-}_2 = 
-m \frac{\spa{\wh4}.6^4}{\spa{\wh4}.5 \spa5.6 \ab{6}{\wh P}{\wh3} \ab{\wh4}{\wh P}{\wh3}
	2\lphatT2.3 p_{4,5,6}^2} \times
\Bigg[ [1\wh3]\ab{3}{\wh P}{2}+\ba{1}{\wh P}{3}[\wh3 2] \Bigg], \hskip50pt
\eeqn
where $D^{--+-}_2$ has $z = -\frac{p_{4,5,6}^2}{\ab{3}{5+6}{4}}$ and
\beqn \label{mmpmshift2}
	\ab{3}{\wh P}{b} &=& \spa3.4[4b]+\spa3.5[5b]+\spa3.6[6b].
\eeqn
For the third diagram we have 
\beqn
\lefteqn{\mathcal D^{--+-}_3 = 
-m\frac{\spa3.5+\frac{2\lp1.6}{\ab{6}{1}{\wh P}}\spa3.6}
	{2\lphatT2.3 p_{3,4,5}^4 [\wh P 6] 2\lp1.6 \spa5.4} \times
\frac{\spa3.4^3}{\spa3.5}  } \\ & & \times \nn
\Bigg[
m\spRb1.6 \frac{p_{3,4,5}^2}{\spa3.5} [\wh P 2]-[1 \wh P] \ab{6}{1}{\wh3}[\wh P 2]+
		[1 \wh P]\ab{6}{1}{\wh P}[\wh3 2]
\Bigg], \hskip171pt
\eeqn
where in the third diagram $z = -\frac{\spa4.5}{\spa3.5}$ and 
$\bra{\wh P}=\bra5+\frac{\spa3.4}{\spa3.5}\bra4$.
%
%

The ``$\spRa4.3$" shift yields the final single-positive-helicity gluon 
amplitude at six points,
\beqn
	A_6\left(\bar{1}^{-}_q,2^{-}_q,3^{-}_g,4^{-}_g,5^{-}_g,6^{+}_g\right) = 
	\mathcal D^{---+}_1+\mathcal D^{---+}_2+\mathcal D^{---+}_3,
\eeqn
where $\mathcal D^{---+}_3$ vanishes.  
For the remaining diagrams, $\mathcal D^{---+}_1$ and $\mathcal D^{---+}_2$, we have
\beqn
\lefteqn{\mathcal D^{---+}_1 = 
-\frac{
\ba{6}{1}{5}+\frac{2\lphat4}{\ba{5}{\wh P}{\wh4}}\ba{6}{1}{\wh4}
}
{2\lp1.6 2\lphat4 \left(2\lp5.6+\frac{2\lphat4}{\ba{5}{\wh P}{\wh4}} \ba{5}{6}{\wh4} \right)
	[45] [34] \left(-2\lp2.3\right)
}   } \label{mmmp1}
\\ & & \nn
\times
\Bigg[
\Big([16] 2\lp1.6 + m\Big(\spRb1.5+\frac{2\lphat4}{\ba{5}{\wh P}{\wh4}}\spRb1.{\wh4}\Big)[65] \Big)
\left(2\lphat4 \spRa3.2+m\spa4.3 [42] \right) \\ &&
 \hskip .5 cm \null +
\Big( \Big(\spRb1.5+\frac{2\lphat4}{\ba{5}{\wh P}{\wh4}}\spRb1.{\wh4}\Big) \ba{6}{\wh P}{\wh4} +
	m[16] \spa5.{\wh4} \Big) \Big(\ba{5}{2}{3} [42] + m[54] \spRa3.2 \Big)
\Bigg] \hskip 72pt  \nn \\ && \null + \nn
m\frac{\spa{\wh4}.5^3}{\spa6.5  (p_1+p_2-p_{\wh3})^4 
	(2\lphat4+\frac{\spa5.6}{\spa{\wh4}.6}\ab{\wh4}{\wh P}{5}) [34] 2\lp2.3} \times \\ && \nn
\Bigg[\Big([14]+\frac{\spa5.6}{\spa{\wh4}.6}[15]\Big)
	\Big( (\ab{3}{2}{6}+\frac{\spa{\wh4}.5}{\spa{\wh4}.6}\ab{3}{2}{5})[42]+
	m([64]+\frac{\spa{\wh4}.5}{\spa{\wh4}.6}[54])\spRa3.2\Big) \\ && \nn 
\hskip .5 cm \null -
	\Big([16]+\frac{\spa{\wh4}.5}{\spa{\wh4}.6}[15]\Big)
	\Big( (\ab{3}{2}{4}+\frac{\spa5.6}{\spa{\wh4}.6}\ab{3}{2}{5})[42]+
	m\frac{\spa5.6}{\spa{\wh4}.6}\spRa3.2\Big)
\Bigg]
\eeqn
and the second diagram 
\beqn
\label{mmmp2}
\mathcal D^{---+}_2 = 
-m \frac{\spa{\wh4}.5^3}{\spa5.6 \ab{6}{\wh P}{\wh3} \ab{\wh4}{\wh P}{\wh3}
	2\lphatT2.3 p_{4,5,6}^2} \times
\Bigg[ [1\wh3]\ab{3}{\wh P}{2}+\ba{1}{\wh P}{3}[\wh3 2] \Bigg].	\hskip73pt
\eeqn
The shifted spinors in the first and second diagrams, (\ref{mmmp1}) and (\ref{mmmp2}), are identical 
to those found in $A_6\left(\bar{1}^{-}_q,2^{-}_q,3^{-}_g,4^{-}_g,5^{+}_g,6^{-}_g\right)$, 
eqs.~(\ref{mmpmshift1}) and~(\ref{mmpmshift2}) respectively.

Finally we write the amplitudes with two negative-helicity gluons, beginning with
the amplitude
\beqn
A_6\left(\bar{1}^{-}_q,2^{-}_q,3^{-}_g,4^{-}_g,5^{+}_g,6^{+}_g\right) = 
\mathcal D^{--++}_1 + \mathcal D^{--++}_2 + \mathcal D^{--++}_3,
\eeqn
where the ``$\spRa3.4$" shift is chosen.  The first diagram, $\mathcal D^{--++}_1$, is 
\beqn
\lefteqn{\mathcal D^{--++}_1 =
\frac
{\left(\ab{4}{\wh P}{5}+\frac{2\lp1.6}{\ab{5}{1}{6}}\ab{4}{\wh P}{6} \right)}
{2\lphat4 \left(2\lphatS4.5+\frac{2\lp1.6}{\ab{5}{1}{6}}\ab{5}{\wh4}{6}\right)
	2\lp1.6 \spa6.5 \left(-2\lp2.3\right) [34] } }
\\ && \times \nn
\Bigg[
\Big(-\spRb1.5\ab{4}{1}{6}+m[16]\spa5.4\Big) \\ & & \nn
\hskip .5 cm \times
\Bigg( 
\ab{\wh3}{2}{5}[\wh4 2]+m [5 \wh4] \spRa{\wh3}.2  + 
\frac{2\lp1.6}{\ab{5}{1}{6}}\left(\ab{\wh3}{2}{6}[\wh4 2]+m[6 \wh4] \spRa{\wh3}.2 \right)
\Bigg) \\&& \nn
\hskip .5 cm \null +
\left(-[16] 2\lphat4+m\spRb1.5[65]\right)
\left( 2\lphat4 \spRa{\wh3}.2 + m\spa4.3 [\wh4 2] \right)
\Bigg]  \hskip 141pt  \\ && \nn 
\null -
\frac{m [56]^3}
	{(p_1+p_2-p_{\wh3})^4 \left(2\lp1.6+\frac{[\wh45]}{[\wh46]}\ab{5}{1}{6}\right)[5\wh4][43] 2\lp2.3}
	\\ && \times \nn
\Bigg[
	\Bigg(
	\Big(\spRb1.4+\frac{[56]}{[\wh46]}\spRb1.5\Big) 
	\Big(\ab{6}{\wh P}{\wh4}+\frac{[\wh45]}{[\wh46]}\ab{5}{\wh P}{\wh4}\Big)  \\ &&  \nn
\hskip .5 cm \null -
	\Big(\spRb1.6+\frac{[\wh45]}{[\wh46]}\spRb1.5\Big)
	\Big(2\lphat4+\frac{[56]}{[\wh46]}\ab{5}{\wh P}{\wh4}\Big)
	\Bigg)  \spRa{\wh3}.2 +
m \spRb1.{\wh3}\frac{p_{\wh4,5,6}^2}{[\wh46]}[\wh42]
\Bigg].
\eeqn
For the diagram $\mathcal D^{--++}_1$ the spinors are shifted with
\beqn
	z &=& \frac{-2\lp2.3}{\ab{4}{2}{3}}, \nn \\ \nn 
	\ab{4}{\wh P}{5} &=& \ab{4}{6-1}{5}, \quad
	\ab{a}{\wh P}{\wh4} = \ab{a}{5+6-1}{\wh4},  \\ 
	\lphat4 &=& \lp1.5 + \lp1.6 - \lp5.6 .
\eeqn
We note that the second term in $\mathcal D^{--++}_1$ arises from a
vanishing term in
$A_5\left(\bar{1}^{-}_q,2^{-}_q,3^{-}_g,4^{+}_g,5^{+}_g\right)$.  This
five-point term is zero for the special quark spins chosen but
contributes a residue at six points since the recursion sums over both
spin states of an internal quark.
The second diagram yields
\beqn
\lefteqn{\mathcal D^{--++}_2 = 
\frac{\ab{\wh3}{2}{5} \spa5.4 + \ab{\wh3}{2}{6} \spa6.4}
{\left(p_1+p_2\right)^2 2\lphatT2.3 p_{4,5,6}^2 } \frac{[65]}{\spa4.5 \spa5.6 [5 \wh4]} 
}
\\ && \nn
\times
\Bigg[
\spRb1.{\wh3} \left(\spa4.5[52]+\spa4.6[62]\right) - 
\left([15]\spa5.4 +[16]\spa6.4 \right) \spRa{\wh3}.2
\Bigg], \hskip 166.7pt
\eeqn
where the $\wh P$ spinors have been eliminated using 
$\spa a.{\wh P} [\wh P b] = \ab{a}{\wh4 +5+6}{b}$, 
and $z = \frac{p_{4,5,6}^2}{\ab{4}{5+6}{3}}$.
For the third diagram we have 
\beqn
\lefteqn{\mathcal D^{--++}_3 = 
-\frac{\left(\ab{\wh3}{2}{5}+\frac{2\lp1.6}{\ab{\wh P}{1}{6}}\ab{\wh3}{2}{6} \right) }
{2\lphatT2.3 \left(2\lphat3+\frac{2\lp1.6}{\ab{\wh P}{1}{6}} \ab{\wh P}{\wh3}{6}\right) 
	2\lp1.6 \spa6.{\wh P} } \times \frac{[35]}{[45][34]} 
} \\ & & \times  \nn
\Bigg[
\left(-\spRb1.{\wh P} \ab{\wh3}{1}{6}+m[16] \spa{\wh P}.{\wh3} \right)
\Big([52]+\frac{2\lp1.6}{\ab{\wh P}{1}{6}}[62]\Big) - 
\left([16] 2\lphatT2.3 - m\spRb1.{\wh P} [65] \right) \spRa{\wh3}.2
\Bigg],
\eeqn
where $z = \frac{[45]}{[35]}$. By using the Schouten identity we find
\beqn
	\bra{\wh 4} &=& \bra4 - \frac{[45]}{[35]} \bra3 = \frac{[34]}{[35]}\bra5 
\nn \\ \wh P &=& \left(\frac{[34]}{[35]} \ang4+\ang5\right) \bbra5.
\eeqn

The next amplitude is 
\beqn
A_6\left(\bar{1}^{-}_q,2^{-}_q,3^{-}_g,4^{+}_g,5^{-}_g,6^{+}_g\right) = 
\mathcal D^{-+-+}_1 + \mathcal D^{-+-+}_2 + \mathcal D^{-+-+}_3
\eeqn
where the shift ``$\spRa6.5$" is chosen to find
\beqn
\lefteqn{\mathcal D^{-+-+}_1 =
\frac{\ab{3}{2}{4}+\frac{2\lphatT1.6}{\ab{\wh P_R}{1}{6}}\ab{3}{2}{6}}
{2\lp2.3 
	\left(2 \lphatRb3 +\frac{2\lphatT1.6}{\ab{\wh P_R}{1}{6}}\ab{\wh P_R}{3}{6}\right)
	2\lphatT1.6  p_{4,5,6}^2 }
\times \frac{[46]^2}{[54][56]}
}
\\ && \nn
 \times
\Bigg[
\left(-\spRb1.{\wh P_R}\ab{3}{1}{6}+m[16]\spa{\wh P_R}.3 \right)
\Big([42]+\frac{2\lphatT1.6}{\ab{\wh P_R}{1}{6}}[62]\Big) \\ && \nn
\hskip .5 cm \null + 
\left(-[16] 2\lp2.3 + m\spRb1.{\wh P_R} [64]\right) \spRa3.2
\Bigg], \hskip 246pt
\eeqn
where $z = \frac{[45]}{[46]}$, so then 
	$\wh P_R = \left(\ang4 + \frac{[56]}{[46]} \ang5\right)\bbra4$ and 
	$2 \lphatRb3 = p_{3,4,\wh5}^2$.
For the second diagram, $\mathcal D^{-+-+}_2$,  we have 
\beqn
\lefteqn{\mathcal D^{-+-+}_2 =
-\frac{\spa3.5^4}{\spa3.4 \spa4.5 \ab{5}{3+4}{6} \ab{3}{4+5}{6}} \times
\frac{1}{p_{3,4,5}^2 \left(p_1+p_2\right)^2 2\lphatT1.6}
} \\ && \times \nn
[6 \wh P]\ab{\wh P}{2}{6} \times 
\Bigg[
\spRb1.{\wh P}[\wh P 6][62]-[16] [6 \wh P]\spRa{\wh P}.2
\Bigg] \hskip 220pt
\eeqn
where $z = \frac{p_{3,4,5}^2}{\ab{5}{3+4}{6}}$ and 
$\ang{\wh P}[\wh P 6] = -\left(p_3+p_4+p_5\right)\bra6$.
The third diagram is 
\beqn
\lefteqn{\mathcal D^{-+-+}_3 = 
\frac{\ab{5}{6-1}{4}}
{2\lphat5 2\lp2.3 \left(2\lphatT4.5+\frac{2\lp2.3}{\ab{3}{2}{4}}\ab{3}{\wh 5}{4}\right)[34]
	\left(-2\lp1.6\right) \spa5.6}
} \\ && \nn \times
\Bigg[
\left([16]\ab{5}{\wh P}{4}+m\spRb1.5[64]\right)
\left(m\spa5.3[42] + \ab{5}{\wh P}{4}\spRa3.2\right) +
\spRb1.5 \ab{5}{\wh P}{6} \ab{3}{2}{4} [42]
\Bigg] \hskip 44pt  \\ && \nn 
\null -
m\frac{\spa3.5^4}{\spa5.6 2\lp1.6 \spa5.4 \spa3.4 (p_1+p_2-p_{\wh6})^4 
	(2\lp2.3+\frac{\spa4.5}{\spa3.5}\ab{3}{2}{4})} \\ && \times \nn
\Bigg[
\big( m \spRb1.5 ([63]+\frac{\spa4.5}{\spa3.5}[64])+
	[16]\big(\ab{5}{6-1}{3}+\frac{\spa4.5}{\spa3.5}\ab{5}{6-1}{4} \big)
\big([\wh5 2]+\frac{\spa3.4}{\spa3.5}[42]\big) 
\\ && \nn 
\hskip .5 cm \null -
\big(m\spRb1.5 ([65]+\frac{\spa3.4}{\spa3.5}[64])+
	[16](2\lphat5+\frac{\spa3.4}{\spa3.5}\ab{5}{6-1}{4})\big)
\big([32]+\frac{\spa4.5}{\spa3.5}[42] \big)
\Bigg] ,
\eeqn
where $z = -\frac{2\lp1.6}{\ab{5}{1}{6}}$. Then 
$\wh P = p_{\wh6} - p_1$ so that $\ab{5}{\wh P}{b} = \ab{5}{6-1}{b}$ and 
$2\lphat5 = -2\lp2.3 -2\lp2.4 + 2\lp3.4$.

The final six-point amplitude with two negative-helicity gluons is found from 
the shift choice $``\spRa5.6"$,
\beqn
A_6\left(\bar{1}^{-}_q,2^{-}_q,3^{-}_g,4^{+}_g,5^{+}_g,6^{-}_g\right) = 
\mathcal D^{-++-}_1 + \mathcal D^{-++-}_2 + \mathcal D^{-++-}_3.
\eeqn
We find for the first diagram in fig.~\ref{4gshift56}
\beqn
\lefteqn{\mathcal D^{-++-}_1 = 
\frac{\spa4.6}{\spa5.6 \spa4.5} \times 
\frac{\ab{6}{1}{\wh P_R}}
{2\lphatT1.6 2\lp2.3 \left(2\lphatR6+\frac{2\lp2.3}{\ab{3}{2}{\wh P_R}}\ab{3}{\wh6}{\wh P_R}\right)	[3 \wh P_R]} } \\ & & \times \nn
\Bigg[
m[1 \wh P_R]\spa6.3[\wh P_R 2] + \spRb1.6\ab{3}{2}{\wh P_R}[\wh P_R 2]-[1 \wh P_R]\ab{6}{1}{\wh P_R}\spRa3.2
\Bigg] \hskip 143pt 
\\ & & \nn 
\null -
m \frac{\spa3.6^4}
	{\spa3.4 \spa4.5 \spa5.6 (p_1+p_2)^4 (2\lp2.3+\frac{\spa4.6}{\spa3.6}\ab{3}{2}{\wh P_R})}
\\ && \nn \times
\Bigg[([13]+\frac{\spa4.6}{\spa3.6}[1 \wh P_R])([\wh6 2]+\frac{\spa3.4}{\spa3.6}[\wh P_R 2])  
-
([1 \wh6]+\frac{\spa3.4}{\spa3.6}[1 \wh P_R])([32]+\frac{\spa4.6}{\spa3.6}[\wh P_R 2]) 
\Bigg], 
\eeqn
where $z = -\frac{\spa4.5}{\spa4.6}$ and  
$\wh P_R = \ang4 \left(\bbra4 + \frac{\spa5.6}{\spa4.6}\bbra5\right)$.
The second diagram is 
\beqn
\lefteqn{
\mathcal D^{-++-}_2 = 
-\frac{[45]^3}{\ab{6}{\wh P}{5} \ab{6}{\wh P}{3} [34]} \times 
\frac{1}{p_{3,4,5}^2  2\lphat6 2\lphatT1.6}  }  \\ && \times \nn
\ab{6}{2}{\wh P} \spa{\wh P}.6 \times 
\Bigg[
-[1 \wh P] \spa{\wh P}.6 \spRa6.2 + \spRb1.6 \spa6.{\wh P} [\wh P 2]
\Bigg], \hskip 193pt
\eeqn
where $z = -\frac{p_{3,4,5}^2}{\ab{6}{3+4}{5}}$, $-2\lphat6 = \left(p_1+p_2\right)^2$,
and $\bra{\wh P} \spa{\wh P}.6 = -\left(p_3+p_4+p_5\right)\ang6$.
The third diagram, $\mathcal D^{-++-}_3$, is 
\beqn
\lefteqn{
\mathcal D^{-++-}_3 = 
\frac{\ab{3}{2}{4}+\frac{2\lphat5}{\ab{4}{\wh P}{5}}\ab{3}{2}{5}}
{2\lp2.3 \left(2\lp3.4+\frac{2\lphat5}{\ab{4}{\wh P}{5}}\ab{4}{3}{5}\right)
	2\lphat5 \spa{\wh5}.4 [65] 2\lp1.6}  } \\ && \times \nn
\Bigg[
\spRb1.6 \ab{4}{\wh P}{5} 
   \left(\ab{3}{\wh P}{5} \bigg([42]+\frac{2\lphat5}{\ab{4}{\wh P}{5}}[52]\bigg) 
	+m[54]\spRa3.2 \right) \\ & & \nn
\hskip .5 cm \null +
[15] \Bigg( \ab{6}{\wh P}{5} 
	\left(m \spa4.3 \bigg([42]+\frac{2\lphat5}{\ab{4}{\wh P}{5}}[52]\bigg)
	-2\lp2.3\spRa3.2 \right) \\ & & \nn
\hskip .5 cm \null +
m \spa6.4 \left(\ab{3}{\wh P}{5} \bigg([42]+\frac{2\lphat5}{\ab{4}{\wh P}{5}}[52]\bigg)+
	m[54]\spRa3.2 \right) \Bigg)
\Bigg] \hskip 151pt \\ \nn && 
\null-
\frac{m [45]^3}{(p_1+p_2-p_{\wh6})^4 
	\left(-2\lphat5-\frac{[34]}{[35]}\ab{4}{\wh P}{5}\right)[43][56] 2\lp1.6} \\ && \nn \times
\Bigg[
\spRb1.6 \Bigg(
	\Big( \ab{3}{\wh P}{5}+\frac{[45]}{[35]}\ab{4}{\wh P}{5} \Big)
	\Big( \spRa{\wh5}.2+\frac{[34]}{[35]}\spRa4.2 \Big) \\ && \nn 
\hskip .5 cm \null -
	\Big( 2\lphat5+\frac{[34]}{[35]}\ab{4}{\wh P}{5} \Big)
	\Big( \spRa3.2+\frac{[45]}{[35]}\spRa4.2 \Big) \Bigg) +
m[15] \frac{p_{3,4,\wh5}^2}{[35]} \spRa6.2
\Bigg],
\eeqn
where $z = \frac{2\lp1.6}{\ab{6}{1}{5}}$, $2\lphat5 =
\left(p_2-p_3-p_4\right)^2$, and $\ab{a}{\wh P}{5} = \ab{a}{6-1}{5}$.
%

\section{Conclusions}

We used the BCFW formalism to calculate the scattering amplitudes with
a pair of massive fermions and four or fewer gluons by shifting
adjacent pairs of massless gluons.  All the helicity configurations
are computed in this way. By deriving an explicit formula for the
shifted spinors of an internal gluon,  no subtleties are 
encountered~\cite{Stirling} with the massive-spinor inner products.

We have performed a number of checks on our results. By performing a
variety of shifts we numerically confirmed that the residues
corresponding to all possible channels are correct.  We note that
shifts where the amplitude is not well behaved for large $z$ can be
used for the purpose of checking a putative amplitude, even though an
on-shell recursion constructed from such a shift would drop terms
coming from the additional large $z$ contribution. To perform the
check we shift a different pair of spinors than the ones used to
construct the amplitude via on-shell recursion and extract each residue
which we then compare to the corresponding BCFW diagram.
This verifies that the answers are consistent, so that we obtain the proper 
residues no matter which
pair of external gluons are shifted.
This check is equivalent to checking all factorization limits of
the amplitude.

We have also checked that our results agree with the amplitudes
presented in ref.~\cite{Schwinnborn}, where formulas for the
amplitudes with identical or all but one identical helicity gluons
were derived from massive scalar amplitudes via supersymmetric Ward
identities.  The five-point amplitudes match the results of
ref.~\cite{Stirling}, which were calculated using BCFW recursion and
Feynman diagrams for certain amplitudes.

The amplitudes presented in this paper are a subset of those
needed for heavy quark plus two-jet production at the LHC.
The calculations of this paper confirm that the BCFW recursion
relations are an effective way to compute amplitudes with massive
quarks. 

\section*{Acknowledgements}

The author thanks Zvi Bern for many helpful discussions and 
Christian Schwinn and Stefan Weinzierl for helpful communications
in comparing our results.  The author also thanks Academic Technology
Services at UCLA for computer support.  


\begingroup\raggedright\endgroup

\end{document}